\documentclass{article}
\usepackage{spconf,amsmath,graphicx,hyperref}

\usepackage{cite}
\usepackage{amsmath,amssymb,amsfonts}
\usepackage{algorithmic}
\usepackage{graphicx}
\usepackage{textcomp}
\usepackage{xcolor}

\usepackage{booktabs}
\usepackage{booktabs}
\usepackage{threeparttable}
\usepackage{epstopdf}

\usepackage{CJKutf8}
\usepackage{multirow}
\usepackage{epstopdf}
\usepackage{bm}
\usepackage{setspace}
\usepackage{mathrsfs}
\usepackage{times}
\usepackage{latexsym}
\usepackage{stfloats}
\usepackage{cases}
\usepackage{array}
\usepackage{fancyhdr}
\usepackage{subfigure}
\usepackage{balance}
\usepackage{algorithm}
\usepackage{mathtools}
\usepackage{booktabs}
\usepackage{float}
\usepackage{threeparttable}
\usepackage{subfigure}
\usepackage{tikz}
\usepackage{color,xcolor}
\usepackage{amssymb}
\usepackage{footnote}

\title{State--Flow Coordinated Representation for MI--EEG Decoding}
\name{Guoqing Cai$^1$   \ \ \ \ \ \ \ \ \ \ 
Shoulin Huang$^3$  \ \ \ \ \ \ \ \ \ \  
Ting Ma$^{1, 2}$ $^{\ast}$ 
\thanks{*Corresponding author} 
}
\address{
{$^1$Harbin Institute of Technology (Shenzhen), Shenzhen, China}\\
{$^2$Peng Cheng Laboratory, Shenzhen, China}\\
{$^3$Guangxi Normal University, Guilin, China}
}

\begin{document}
\ninept
\maketitle

\begin{abstract}
Motor Imagery (MI) Electroencephalography (EEG) signals contain two crucial and complementary types of information: 
state information, which captures the global context of the task, and flow information, which captures fine-grained temporal dynamics.
However, existing deep decoding models typically focus on only one of these information streams, resulting in unstable learning and sub-optimal performance. 
To address this, we propose the State–Flow Coordinated Network (StaFlowNet), a novel architecture that explicitly separates and coordinates state and flow information. 
We first employ a dual-branch design to extract the global state vector and temporal flow features separately. 
Critically, a novel state-modulated flow module is proposed to dynamically refine the learning of flow information. 
This modulated mechanism effectively integrates global context with fine-grained dynamics, thereby significantly enhancing task discriminability and decoding performance. 
Experiments on three public MI-EEG datasets demonstrate that StaFlowNet significantly outperforms state-of-the-art methods. 
Ablation studies further confirm that the state-modulated mechanism plays a crucial role in  enhancing feature discriminability and  overall performance.

\end{abstract}


\section{Introduction}

Brain–Computer Interfaces (BCIs) enable direct communication between the brain and external devices, offering new opportunities for rehabilitation and assistance to individuals with motor impairments.
Among BCI paradigms, Motor Imagery (MI) based on Electroencephalography (EEG) is widely used due to its non--invasive nature and ease of deployment \cite{2022ICASSP1}.
The primary objective in MI-EEG based BCIs is to accurately decode a user's motor intention from complex EEG signals. 
Despite significant progress, achieving accurate decoding of users' motor intentions from these signals remains a major challenge.

MI-EEG signals contain two distinct yet complementary types of information \cite{2025MARBLE}. 
The first is the global, quasi-stationary information associated with Event-Related Desynchronization/Synchronization (ERD/ERS)  \cite{ERDERS}.
 We define this as the $ \mathbf{state}$ information.
During MI tasks, it manifests as sustained power modulation in specific frequency bands over the sensorimotor cortex. 
State information captures the macroscopic functional context of the task and forms the foundation for accurate decoding. 
Classical methods, such as Common Spatial Pattern (CSP), target this component by designing spatial filters to maximize inter-class variance \cite{2024MLCSP}. 
While effective, such methods rely on trial-wide statistics (e.g., covariance), which smooth out temporal variations and sacrifice fine-grained dynamics.

In contrast to the stable state information, MI-EEG signals also contain transient and fine-grained temporal dynamics, referred to as the $\mathbf{flow}$  information. 
Flow information includes instantaneous phase relationships, subtle frequency shifts, and the temporal evolution of ERD/ERS, such as onset latency and peak timing \cite{2025ICASSP3}. 
These rapidly changing,  non-stationary features provide rich discriminative cues that are crucial for accurate decoding. 
To capture the flow information, many deep learning approaches have been proposed. 
Models such as Convolutional Neural Networks (CNNs), Recurrent Neural Networks (RNNs), and Transformers 
attempt to learn temporal dependencies directly \cite{2022eegConformer, 2023ICASSP2, 2018eegnet, 2024TransNet, 2024msvtnet, 2023ICASSPGRU, 2025CGNet}. 
However, these end-to-end approaches face a critical bottleneck.
 They attempt to model the temporal flow without an architectural prior that leverages the stable, global context provided by state information. 
 This lack of guidance forces the models to learn from noisy, non-stationary dynamics, making the learning process unreliable and leading to sub-optimal performance. 

To address this, 
we propose the State--Flow Coordinated Network (StaFlowNet), which integrates state and flow processing within  a unified  architecture. 
StaFlowNet adopts a dual-branch design to separately extract state and flow information.
A dedicated State Encoder collapses the temporal dimension into a global state vector that captures the macroscopic context of each trial.
In parallel, a Flow Encoder processes a time-differenced signal to retain fine-grained temporal evolution.
The extracted flow features are then passed through a multi-scale temporal pyramid that hierarchically models dynamic patterns. 
At each pyramid layer, a state-modulated flow module leverages the global state vector to refine the flow representation,
 ensuring that local temporal dynamics are consistently guided by the macroscopic context. 
 Through this coordinated fusion of state and flow, StaFlowNet mitigates the instability of direct temporal modeling and enhances the discriminability of learned features. 
 Experiments on three public MI-EEG datasets demonstrate that StaFlowNet consistently outperforms existing methods. 
 Visualization further reveals that state and flow capture distinct spatial patterns. 
 Moreover, flow features modulated by state information exhibit stronger discriminability than flow alone.

\section{Methodology}

\begin{figure*}
	\centering%
	\includegraphics[scale=0.5,trim=1.1cm 0.5cm 0.1cm 0.5cm,clip]{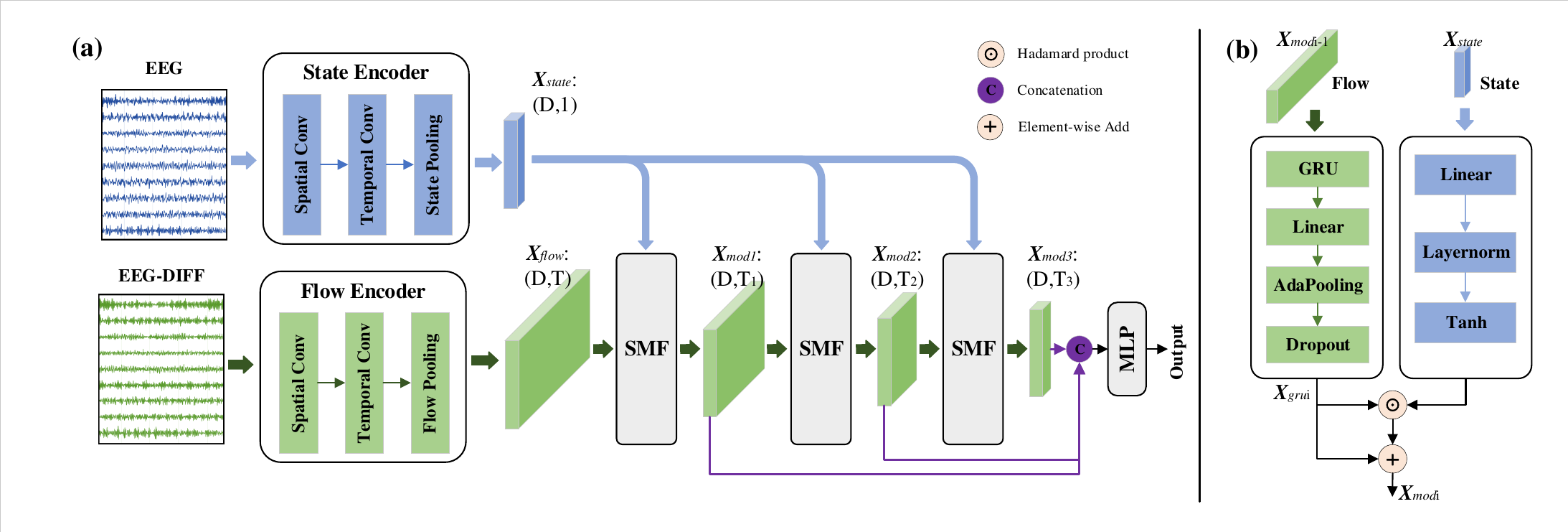}
	\caption{
    (a)  Overview of the proposed StaFlowNet architecture.
          It consists of three key modules: a State Encoder, a Flow Encoder, and a State-Modulated Flow (SMF) module. The fused representation is passed to an MLP for classification. 
    (b) Structure of the SMF module. Each GRU output is adaptively modulated by the state vector through element-wise gating.}
    \label{fig:StaFlowNet}
\end{figure*}

Given a multi-channel EEG trial $\mathbf{X} \in \mathbb{R}^{C \times T_{\text{in}}}$, where $C$ is the number of channels and $T_{\text{in}}$ is the number of time points, 
our goal is to extract both global state information and local flow dynamics. 
To this end, we propose the StaFlowNet architecture, 
which comprises a State Encoder, a Flow Encoder, a State-Modulated Flow (SMF) module, as shown in Fig. \ref{fig:StaFlowNet}. 

\subsection{State Encoder}

The State Encoder is designed to extract macroscopic, quasi-stationary state information that represents the global cognitive context of the trial. 
These features, such as ERD/ERS patterns, are relatively stable and exhibit high discriminative power.
We apply spatiotemporal convolutional operations followed by adaptive average pooling to obtain a compact state representation:
\begin{equation}
    {\mathbf{X}}_{\text{state}} = \text{Dropout}\left( \text{Pool} \left( \sigma \left( \text{BN} \left( \mathbf{W}_2 * (\mathbf{W}_1 * \mathbf{X}) \right) \right) \right) \right)
\end{equation}
where the operator $*$ denotes convolution.
 $\mathbf{W}_1$ is the spatial convolution with kernel size $(C,1)$, and $\mathbf{W}_2$ is the temporal convolution with  kernel size $(1,32)$. 
 $\sigma(\cdot)$ denotes the ELU activation function, and $\text{BN}$ represents batch normalization. 
 $\text{Pool}(\cdot)$ is an adaptive 2D pooling operation that outputs a fixed-size $(1,1)$ tensor. 
 The resulting vector $\mathbf{X}_{\text{state}} \in \mathbb{R}^{D \times 1}$ serves as a compact global representation of the trial's state, with $D = 80$ in our implementation.

\subsection{Flow Encoder}
To emphasize transient dynamics, we compute the temporal derivative of the raw EEG signals instead of absolute amplitude:
\begin{equation}
    \mathbf{X}_{\text{diff}}[:, t] = 
        \begin{cases}
        0, & t = 0, \\
        \mathbf{X}[:, t] - \mathbf{X}[:, t-1], & 1 \leq t < T_{\text{in}},
        \end{cases}
\end{equation}

The differenced signals are then passed through the same convolutional pipeline as the State Encoder, 
consisting of spatial convolutions, temporal convolutions, ELU activation, batch normalization, and dropout. 
The only architectural difference from State Encoder lies in the use of average pooling:
\begin{equation}
   \mathbf{X}_{\text{flow}} = \text{Dropout}\left( \text{Pool} \left( \sigma \left( \text{BN} \left( \mathbf{V}_2 * (\mathbf{V}_1 * \mathbf{X}_{\text{diff}}) \right) \right) \right) \right)
\end{equation}
where $\mathbf{V}_1$ and $\mathbf{V}_2$ denote the spatial and temporal convolution, respectively, 
and $\text{Pool}(\cdot)$ indicates 2D average pooling with a kernel of $(1, 48)$ and  a stride of $(1, 32)$. 
The output $\mathbf{X}_{\text{flow}} \in \mathbb{R}^{D \times T}$ forms the flow information, preserving the  fine-grained temporal evolution across $T$ time steps.

\subsection{State-Modulated Flow Module (SMF)}

The SMF module is the core of StaFlowNet. 
It aims to extract temporal features at multiple time scales and modulate them using the state vector $\mathbf{X}_{\text{state}} $.

\subsubsection{Temporal GRU Encoding}
We employ a three-level bidirectional GRU pyramid to capture temporal dependencies at multiple resolutions \cite{2023ICASSPGRU}. 
At the $i$-th level, the flow feature is first transposed to match the temporal input shape 
and then processed by a bidirectional GRU followed by a linear projection to reduce the feature dimension to $D$, as illustrated in Fig. \ref{fig:StaFlowNet}(b). 
\begin{equation}
  \mathbf{X}_{\text{gru}i} = \text{Pool}_{T_i}\left( \text{Linear}\left( \text{GRU}_i(\mathbf{X}_{\text{mod}i-1} ^{\top}) \right) \right) \in \mathbb{R}^{D \times T_i}
\end{equation}
where $\mathbf{X}_{\text{mod}i-1}$ denotes the input from the previous level (or the initial flow features $  \mathbf{X}_{\text{flow}} $ for $i=1$). 
$\text{Pool}_{T_i}(\cdot)$ denotes adaptive average pooling into a fixed length $T_i$.
 This hierarchical design allows the model to extract temporally compressed representations across different scales ($T_1=16$, $T_2=4$, $T_3=1$), 
 thereby enhancing its capacity to model both short- and long-range dynamics.

\subsubsection{State-Based Modulation}
To incorporate state-based modulation, each temporal feature $\mathbf{X}_{\text{gru}i}$ is refined using the global state vector $\mathbf{X}_{\text{state}}$.
 The modulation is computed as:
\begin{equation}
\mathbf{m} = \tanh\left( \text{LN} \left( \mathbf{W}_m \mathbf{X}_{\text{state}} \right) \right) \in \mathbb{R}^{D \times 1},
\end{equation}
\begin{equation}
\mathbf{X}_{\text{mod}i} = \mathbf{X}_{\text{gru}i} \odot (1 + \mathbf{m})  \in \mathbb{R}^{D \times T_i},
\end{equation}
where $\mathbf{W}_m \in \mathbb{R}^{D \times D}$ is a learnable weight matrix, $\text{LN}(\cdot)$ denotes layer normalization, and $\odot$ represents element-wise multiplication.

\subsubsection{Feature Fusion}

The modulated outputs from different levels are concatenated to form the final spatiotemporal representation:
\begin{equation}
    \mathbf{Z} = \text{Concat}(\mathbf{X}_{\text{mod}1}, \mathbf{X}_{\text{mod}2}, \mathbf{X}_{\text{mod}3}) \in \mathbb{R}^{D \times T_{\text{sum}}} 
\end{equation}
where $T_{\text{sum}}=T_1 + T_2 + T_3 = 21 $. This hierarchical fusion strategy allows StaFlowNet to aggregate multi-scale temporal cues under state-based modulation.

\subsection{Classification Head}

The final representation $\mathbf{Z}$ is flattened and passed to a multi-layer perceptron (MLP) for classification:
\begin{equation}
\hat{\mathbf{y}} = \text{MLP}(\text{Flatten}(\mathbf{Z})) \in \mathbb{R}^{C_{\text{cls}}},
\end{equation}
where $C_{\text{cls}}$ is the number of classes. The MLP contains two hidden layers with ELU activation, batch normalization, and dropout.

\section{Experiments}

\subsection{Datasets} 
We evaluate our model on three public MI-EEG datasets:

BCI Competition IV-2a (BCI-IV 2a) \cite{2000FBCSP}:
This dataset contains EEG recordings from 9 subjects performing four MI tasks.
 Signals were recorded from 22 EEG channels at 250 Hz. Each subject completed two sessions on separate days, with 288 trials per session.

BCI Competition IV-2b (BCI-IV 2b) \cite{2000FBCSP}:
This dataset includes EEG data from 9 subjects performing two MI tasks (left-- and right-- hand) within five sessions on separate days. 
Each session contains 120 trials. 
Recordings were made using 3 EEG channels (C3, Cz, C4) and 3 EOG channels at 250 Hz.
In this study, we use only the EEG channels.

OpenBMI \cite{2019OpenBMI}:
A large-scale EEG dataset collected from 54 subjects. We use the subset corresponding to the binary MI task (left-- and right-- hand). EEG was recorded using 62 channels at 1000 Hz and downsampled to 250 Hz for consistency. 
Each subject participated in two recording sessions, with 200 trials per session.

We followed a consistent preprocessing procedure across datasets.
For each trial, EEG segments were extracted from 0 to 4 seconds after the cue onset. 
A 5th-order Butterworth bandpass filter with a range of 4–40 Hz was applied to the data.

\subsection{Comparison Methods}
We compare StaFlowNet with a diverse set of representative deep learning baselines covering compact CNNs, and Transformer-based architectures. 
EEGNet is a compact and widely adopted CNN baseline for EEG-based BCI applications \cite{2018eegnet}. 
ShallowNet is a simple yet strong baseline that has been widely used in MI-EEG decoding \cite{2017deepnet}. 
FBCNet is a representative state-oriented model that embeds the FBCSP pipeline into an end-to-end deep network \cite{2021fbcnet}. 
LightConvNet further improves FBCNet by enhancing temporal modeling within the same FBCSP-style design \cite{2023LightConvNet}. 
We also include three Transformer-based methods, EEGConformer, TransNet, and MSVTNet, which model temporal dependencies in EEG signals with different attention and sequence modeling strategies \cite{2022eegConformer,2024TransNet,2024msvtnet}. 
These baselines provide complementary coverage of state-focused and sequence-focused paradigms in EEG deep learning.

\subsection{Implementation Details}
In this study, all models were trained with the cross-entropy loss and optimized using Adam with a learning rate of 0.001.
Training was capped at 1000 epochs. We used early stopping with patience = 100 to mitigate overfitting \cite{2022eegConformer}.
We followed a cross-session evaluation protocol for all datasets. For BCI-IV 2a and OpenBMI, the first session was used for training and the second for testing. For BCI-IV 2b, the first three sessions were used for training and the last two for testing.
To reduce the effect of random initialization, each experiment was repeated 10 times with different random seeds. 
For each subject, we first averaged the accuracy over the 10 runs, resulting in one accuracy value per subject for each method. 
We then performed the paired Wilcoxon signed-rank test across subjects to compare two methods under the same protocol \cite{2024msvtnet}.
\section{Results and Discussion}

\subsection{Performance Comparison}

\begin{table*}[htbp]
\footnotesize
\centering
\renewcommand\arraystretch{1}
\tabcolsep=0.2cm 
\caption{Comprehensive performance comparison on the three datasets. The best result is highlighted in \textbf{bold}, and the second-best is \underline{underlined}.  
The $p$-value indicates the  Wilcoxon signed-rank test comparing each method's accuracy to StaFlowNet.}
\label{tab:results_all_datasets}
\begin{tabular}{l|lcc|lcc|lcc}
    \toprule
    & \multicolumn{3}{c|}{\textbf{BCI-IV 2a Dataset}} & \multicolumn{3}{c|}{\textbf{BCI-IV 2b Dataset}}  & \multicolumn{3}{c}{\textbf{OpenBMI Dataset}} \\
    \cmidrule(lr){2-4} \cmidrule(lr){5-7} \cmidrule(lr){8-10}
    Methods & Accuracy (\%) & Kappa & f1-score & Accuracy (\%) & Kappa & f1-score & Accuracy (\%) & Kappa & f1-score \\
    \midrule
    EEGNet       & 69.10 $\pm$ 14.04 $^{**}$ & 0.5879 & 0.6803          & 77.56 $\pm$ 12.42 & 0.5513 & 0.7750         & 70.87 $\pm$ 13.69 $^{***}$ & 0.4175 & 0.7087 \\
    ShallowNet   & 72.49 $\pm$ 12.35 $^{**}$ & 0.6332 & 0.7144          & 72.26 $\pm$ 13.02 $^{**}$ & 0.4452 & 0.7218         & 74.99 $\pm$ 13.01 $^{***}$ & 0.4998 & 0.7464 \\
    FBCNet       & 77.57 $\pm$ 10.55 $^{**}$ & 0.7009 & 0.7658         & 70.25 $\pm$ 12.96 $^{**}$ & 0.4051 & 0.6953          & 74.74 $\pm$ 14.24 $^{***}$ & 0.4948 & 0.7389 \\
    EEGConformer & 78.41 $\pm$ 12.09 $^{**}$ & 0.7122 & 0.7811          & \underline{77.96 $\pm$ 12.81} $^{*}$& \underline{0.5591} & \underline{0.7784}           & \underline{77.91 $\pm$ 13.43} $^{**}$ & \underline{0.5581} & \underline{0.7791} \\
    LightConvNet & 76.54 $\pm$ 13.23 $^{**}$ & 0.6872 & 0.7539         & 72.95 $\pm$ 14.36 $^{**}$ & 0.4589 & 0.7235           & 73.62 $\pm$ 16.42 $^{***}$& 0.4725 & 0.7362 \\
    TransNet     & 76.18 $\pm$ 11.78 $^{**}$ & 0.6824 & 0.7495          & 76.73 $\pm$ 13.67 $^{**}$ & 0.5345 & 0.7664          & 76.55 $\pm$ 15.46 $^{**}$& 0.5309 & 0.7621 \\
    MSVTNet      & \underline{78.99 $\pm$ 12.21} $^{**}$ & \underline{0.7198} & \underline{0.7846} &       74.98 $\pm$ 14.65 $^{**}$  & 0.4996 & 0.7473           & 77.64 $\pm$ 14.89 $^{*}$& 0.5528 & 0.7754 \\
    StaFlowNet   & \textbf{80.14 $\pm$ 12.02} & \textbf{0.7351} & \textbf{0.7957} & \textbf{79.02 $\pm$ 12.77} & \textbf{0.5804} & \textbf{0.7864} & \textbf{79.51 $\pm$ 12.97} & \textbf{0.5902} & \textbf{0.7915} \\
    \bottomrule
    
    \multicolumn{10}{l}{$^{***}$ indicates $p < 0.001$, $^{**}$ indicates $p < 0.01$, $^{*}$ indicates $p < 0.05$.}
    
\end{tabular}
\end{table*}

Table~\ref{tab:results_all_datasets} reports the performance of StaFlowNet and all baselines on three MI-EEG datasets, 
measured by classification accuracy, Cohen's Kappa, and F1-score. 
Accuracy is reported as mean $\pm$ standard deviation. 
The results clearly indicate that StaFlowNet consistently achieves the best performance across all three datasets. 
On the BCI-IV 2a dataset, StaFlowNet obtains a mean accuracy of 80.14\%, outperforming the second-best method, MSVTNet (78.99\%), by a margin of 1.15\%. 
On the BCI-IV 2b dataset, it achieves 79.02\% accuracy, surpassing the runner-up, EEGConformer (77.96\%). 
Similarly, on the large-scale OpenBMI dataset, StaFlowNet leads with 79.51\% accuracy, again outperforming EEGConformer (77.91\%). 
The superior Kappa and F1-scores further corroborate the robustness and balance of our model's predictions.
The statistical analysis shows that StaFlowNet's improvements are statistically significant over nearly all competing methods.

\subsection{Ablation Study} 

\begin{table}[htbp]
	\footnotesize
	\caption{Ablation results for accuracy on the three datasets.  The best result is highlighted in \textbf{bold}.}
	\label{tab:ablation}
	\centering
	\renewcommand\arraystretch{1}
	\begin{tabular}{l|l|l|l}
		\toprule
		Methods & \textbf{BCI-IV 2a} & \textbf{BCI-IV 2b} & \textbf{OpenBMI} \\
		\midrule
		StateOnly & 78.16 $^{**}$ & 76.42 $^{**}$ & 78.12 $^{**}$\\
		FlowOnly & 79.14 $^{*}$ & 77.58 $^{*}$ & 78.20 $^{**}$\\
		RandomState & 78.98 $^{*}$ & 77.64 $^{*}$ & 77.90 $^{**}$\\
		Concat & 79.13 $^{*}$ & 78.10 $^{*}$ & 77.81 $^{**}$ \\
		StaFlowNet & \textbf{80.14} & \textbf{79.02} & \textbf{79.51} \\
		\bottomrule
		\multicolumn{4}{l}{ $^{**}$ indicates $p < 0.01$, $^{*}$ indicates $p < 0.05$.}
	\end{tabular}
\end{table}

To validate the effectiveness of the core components in StaFlowNet, we conducted a comprehensive ablation study. 
We designed four variants of our model.
StateOnly: Uses only the State Encoder branch and the classifier.
FlowOnly: Uses only the Flow Encoder and the temporal pyramid, but without any state-based modulation.
RandomState: The full StaFlowNet architecture, but the learned state vector is replaced by a random vector.
Concat: A simplified fusion model where the state vector and the final flow representation are simply concatenated before the classifier, bypassing our modulation mechanism.

The results, summarized in Table \ref{tab:ablation}, provide several key insights. First, FlowOnly consistently outperforms StateOnly, 
suggesting that the temporal flow dynamics contain richer discriminative information than the static state alone. 
Second, our StaFlowNet significantly outperforms FlowOnly, demonstrating that the introduction of state guidance provides a substantial performance boost. 
Third, the RandomState variant performs similarly to FlowOnly and markedly worse than StaFlowNet. 
This is a crucial finding, confirming that the performance gain comes from the learned state information, not merely from the architectural complexity. 
Finally, StaFlowNet also surpasses the Concat model, proving that our proposed state-based modulation is a more effective fusion strategy than feature concatenation. 
Collectively, these results validate our design choices, highlighting the necessity of both state and flow information streams.

\subsection{Visualization}
\begin{figure}
	\centering%
	\includegraphics[scale=0.5,trim=0.1cm 1.2cm 0.1cm 0.5cm,clip]{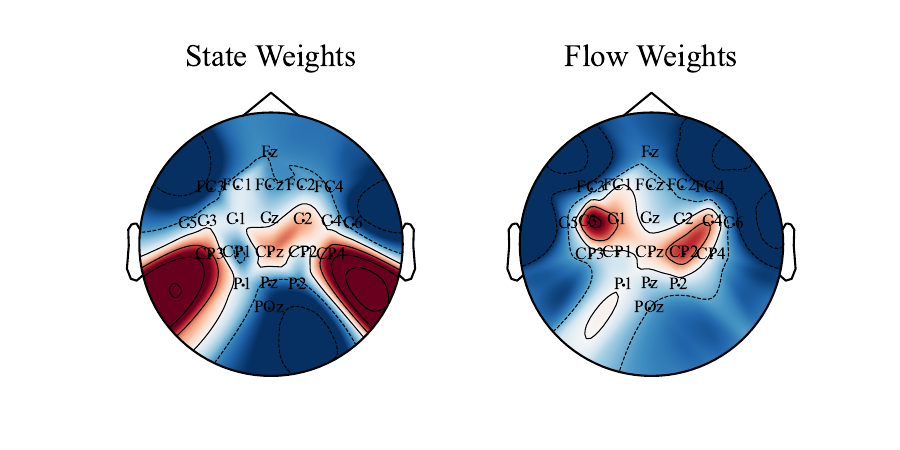}
	\caption{Visualization of learned spatial weights from the State Encoder and Flow Encoder on subject 7 (BCI-IV 2a). }
    \label{fig:topplot}
\end{figure}

\begin{figure}[htbp]
	\centering
	\subfigure[Feature distributions from StaFlowNet]{\includegraphics[scale=0.27,trim=0.1cm 0.3cm 0cm 0.9cm,clip]{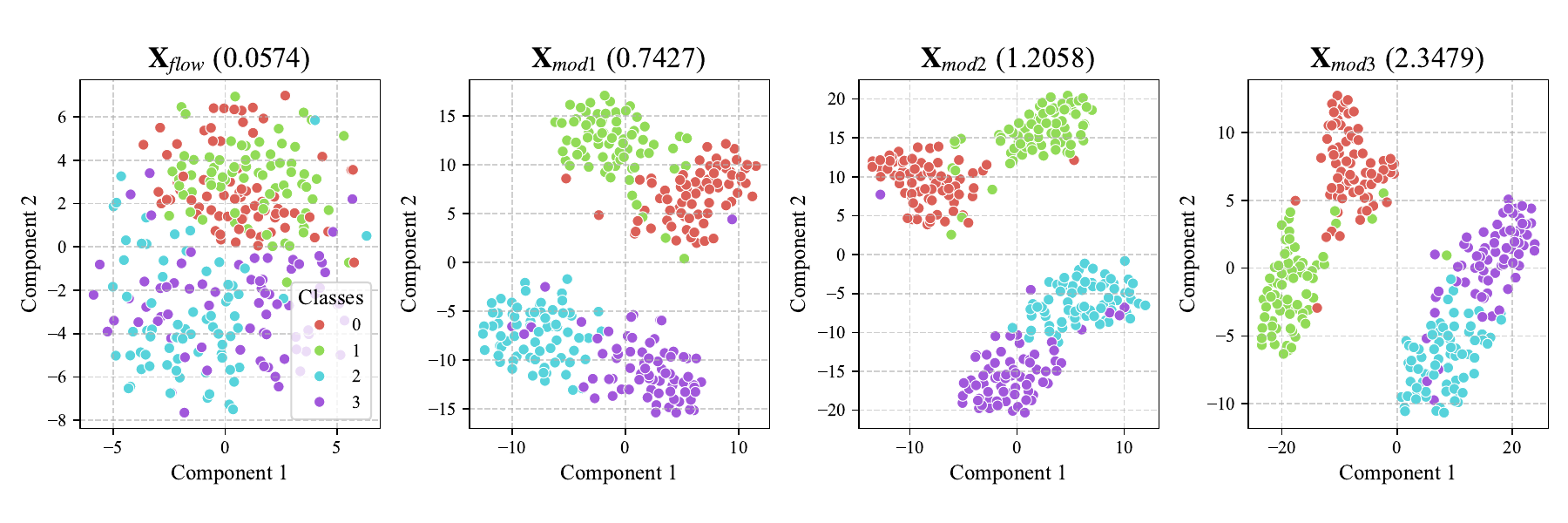}}
	\subfigure[Feature distributions from the FlowOnly variant ]{\includegraphics[scale=0.27,trim=0.1cm 0.3cm 0cm 0.9cm,clip]{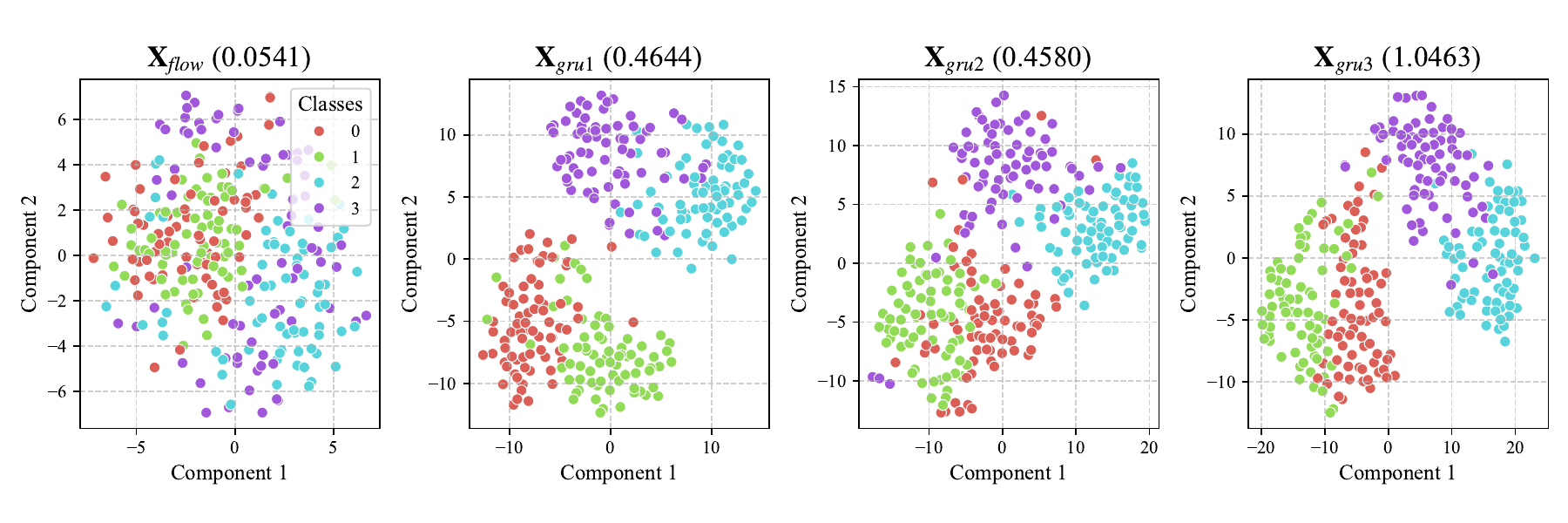}}
	\caption{Feature distributions (t-SNE) from the test set of subject 7 (BCI-IV 2a) at successive network layers. 
		Fisher scores (in parentheses) are computed at each stage to quantify class separability, with higher scores indicating greater inter-class distance relative to intra-class variance.}
	\label{fig:total}
\end{figure}

To better understand the functions of StaFlowNet, we conducted two visualization analyses.
First, we visualized the learned spatial weights from the spatial convolutional layers of the State and Flow encoders \cite{2025CGNet}, as shown in Fig. \ref{fig:topplot}. 
The weights exhibit clear differences, indicating that the two branches focus on distinct aspects of MI-EEG signals.  
This supports our design assumption that state and flow may carry complementary information.
Second, to evaluate the impact of the state-modulated mechanism, we compared feature evolution between StaFlowNet and its FlowOnly variant (without state guidance). 
Fig. \ref{fig:total} presents this comparison on test data from Subject 7 in the BCI-IV 2a dataset. 
In the FlowOnly model (Fig. \ref{fig:total}b), class separation improves gradually across GRU layers, but remains limited. 
In contrast, StaFlowNet (Fig. \ref{fig:total}a) exhibits substantially enhanced feature separability at each pyramid level. 
Besides, the steadily increasing Fisher scores in StaFlowNet further confirm that the state-guided modulation actively transforms the feature space into a more discriminative representation.

\section{Conclusion}

In this study, 
we propose StaFlowNet, a novel architecture for MI-EEG decoding that explicitly separates and coordinates two complementary types of neural information: the global state and the dynamic flow. 
The state provides a robust global context, while the flow retains fine-grained temporal variations.
By using a dual-branch design and introducing a state-modulated mechanism over a multi-scale temporal pyramid, 
StaFlowNet effectively integrates global context with local dynamics. 
Extensive experiments on three datasets demonstrate that StaFlowNet consistently outperforms representative deep learning models.
Visualization analyses further confirm that the state and flow branches capture distinct spatial patterns. 
Critically,  the state-modulated mechanism significantly improves feature separability across layers, leading to more discriminative representations.
Our findings underscore the importance of coordinating state and flow representations for MI-EEG analysis and suggest a promising architectural direction for improving MI-EEG decoding performance.

\bibliographystyle{IEEEtran}%
\bibliography{refs}

\end{document}